\documentstyle[amssymb,preprint,aps]{revtex}

\begin{document}
\title{{\normalsize Published in Foundations of Physics Letters, vol. 12, pp. 291-8
(1998)}\\
Maxwell equations as the one-photon quantum equation\\
}
\author{A. Gersten}
\address{Department of Physics, Ben-Gurion University of the Negev, Beer-Sheva, Israel%
\\
e:mail - gersten@bgumail.bgu.ac.il}
\date{May 7, 1999}
\maketitle

\begin{abstract}
Maxwell equations (Faraday and Ampere-Maxwell laws) can be presented as a
three component equation in a way similar to the two component neutrino
equation. However, in this case, the electric and magnetic Gauss's laws can
not be derived from first principles. We have shown how all Maxwell
equations can be derived simultaneously from first principles, similar to
those which have been used to derive the Dirac relativistic electron
equation. We have also shown that equations for massless particles, derived
by Dirac in 1936, lead to the same result.$\ $The complex wave function,
being a linear combination of the electric and magnetic fields, is a locally
measurable and well understood quantity. Therefore Maxwell equations should
be used as a guideline for proper interpretations of quantum theories.
\end{abstract}

\pacs{Key words: Maxwell equations, one photon, quantum equation.}

The Maxwell equations (except for the electric and magnetic Gauss's law) can
be presented by a three component equation in a way similar to the two
component neutrino equation. This was already known to Oppenheimer \cite
{oppenheimer} and to Majorana \cite{mignani}, \cite{gersten}. Also this type
of equation is a particular case of a more general equation for any spin
derived by Weinberg \cite{weinberg}. There is a continuous interest in this
equation even to this day \cite{good}, \cite{tucker}, \cite{ahluwalia}, \cite
{dvoe}. However one of the drawbacks of the above derivations is that the
electric and magnetic Gauss's laws are not derived from first principles.

The aim of the present latter is to complement the above mentioned works,
and to derive all Maxwell equations directly from a decomposition similar to
that which was used to derive the Dirac relativistic electron equation.

The Dirac equation is derived from the relativistic condition on the Energy $%
E,$ mass $m,$ and momentum ${\bf \vec{p}}$:

\begin{equation}
\left( E^{2}-c^{2}{\bf \vec{p}}^{2}-m^{2}c^{4}\right) I^{(4)}\Psi =0,
\label{n1}
\end{equation}
where $I^{(4)}$ is the $4\times 4$ unit matrix and $\Psi $ is a four
component column (bispinor) wave function. Eq. (\ref{n1}) is decomposed into

\begin{equation}
\left[ EI^{(4)}+\left( 
\begin{array}{cc}
mc^{2}I^{(2)} & c{\bf \vec{p}\cdot \vec{\sigma}} \\ 
c{\bf \vec{p}\cdot \vec{\sigma}} & -mc^{2}I^{(2)}
\end{array}
\right) \right] \left[ EI^{(4)}-\left( 
\begin{array}{cc}
mc^{2}I^{\left( 2\right) } & c{\bf \vec{p}}\cdot {\bf \vec{\sigma}} \\ 
c{\bf \vec{p}\cdot \vec{\sigma}} & -mc^{2}I^{\left( 2\right) }
\end{array}
\right) \right] \Psi =0,  \label{n2}
\end{equation}
where $I^{\left( 2\right) }$ is the $2\times 2$ unit matrix and ${\bf \vec{%
\sigma}}$ is the Pauli spin one-half vector matrix with the components

\begin{equation}
\sigma _{x}=\left( 
\begin{array}{cc}
0 & 1 \\ 
1 & 0
\end{array}
\right) ,\quad \sigma _{y}=\left( 
\begin{array}{cc}
0 & -i \\ 
i & 0
\end{array}
\right) ,\quad \sigma _{z}=\left( 
\begin{array}{cc}
1 & 0 \\ 
0 & -1
\end{array}
\right) ,\quad I^{\left( 2\right) }=\left( 
\begin{array}{cc}
1 & 0 \\ 
0 & 1
\end{array}
\right) .  \label{n3}
\end{equation}

The two component neutrino equation can be derived from the decomposition

\begin{equation}
\left( E^{2}-c^{2}{\bf \vec{p}}^{2}\right) I^{\left( 2\right) }\psi =\left[
EI^{\left( 2\right) }-c{\bf \vec{p}\cdot \vec{\sigma}}\right] \left[
EI^{\left( 2\right) }+c{\bf \vec{p}\cdot \vec{\sigma}}\right] \psi =0,
\label{n6}
\end{equation}
where $\psi $ is a two component spinor wavefunction.

We shall derive the photon equation from the following decomposition

\begin{equation}
\left( \frac{E^{2}}{c^{2}}-{\bf \vec{p}}^{2}\right) I^{\left( 3\right) }{\bf %
=}\left( \frac{E}{c}I^{\left( 3\right) }-{\bf \vec{p}\cdot \vec{S}}\right)
\left( \frac{E}{c}I^{\left( 3\right) }+{\bf \vec{p}\cdot \vec{S}}\right)
-\left( 
\begin{array}{ccc}
p_{x}^{2} & p_{x}p_{y} & p_{x}p_{z} \\ 
p_{y}p_{x} & p_{y}^{2} & p_{y}p_{z} \\ 
p_{z}p_{x} & p_{z}p_{y} & p_{z}^{2}
\end{array}
\right) =0,  \label{n7}
\end{equation}
where $I^{\left( 3\right) }$ is a $3\times 3$ unit matrix, and ${\bf \vec{S}}
$ is a spin one vector matrix with components

\begin{equation}
S_{x}=\left( 
\begin{array}{ccc}
0 & 0 & 0 \\ 
0 & 0 & -i \\ 
0 & i & 0
\end{array}
\right) ,\quad S_{y}=\left( 
\begin{array}{ccc}
0 & 0 & i \\ 
0 & 0 & 0 \\ 
-i & 0 & 0
\end{array}
\right) ,\quad S_{z}=\left( 
\begin{array}{ccc}
0 & -i & 0 \\ 
i & 0 & 0 \\ 
0 & 0 & 0
\end{array}
\right) ,\quad I^{\left( 3\right) }=\left( 
\begin{array}{ccc}
1 & 0 & 0 \\ 
0 & 1 & 0 \\ 
0 & 0 & 1
\end{array}
\right) ,  \label{n8}
\end{equation}
and with the properties

\begin{equation}
\left[ S_{x},S_{y}\right] =iS_{z},\quad \left[ S_{z},S_{x}\right]
=iS_{y},\quad \left[ S_{y},S_{z}\right] =iS_{x},\quad {\bf \vec{S}}%
^{2}=2I^{\left( 3\right) }.  \label{n9}
\end{equation}
The decomposition (\ref{n7}) can be verified directly by substitution.

It will be crucial to note that the matrix on the right hand side of Eq. (%
\ref{n7}) can be rewritten as:

\begin{equation}
\left( 
\begin{array}{ccc}
p_{x}^{2} & p_{x}p_{y} & p_{x}p_{z} \\ 
p_{y}p_{x} & p_{y}^{2} & p_{y}p_{z} \\ 
p_{z}p_{x} & p_{z}p_{y} & p_{z}^{2}
\end{array}
\right) =\left( 
\begin{array}{c}
p_{x} \\ 
p_{y} \\ 
p_{z}
\end{array}
\right) \left( 
\begin{array}{ccc}
p_{x} & p_{y} & p_{z}
\end{array}
\right) .  \label{n10}
\end{equation}
From Eqs. $\left( \ref{n7}-\ref{n8}\right) $ and \ $\left( \ref{n10}\right) $%
, the photon equation can be obtained form

\begin{equation}
\left( \frac{E^{2}}{c^{2}}-{\bf \vec{p}}^{2}\right) {\bf \vec{\Psi}=}\left( 
\frac{E}{c}I^{\left( 3\right) }-{\bf \vec{p}\cdot \vec{S}}\right) \left( 
\frac{E}{c}I^{\left( 3\right) }+{\bf \vec{p}\cdot \vec{S}}\right) {\bf \vec{%
\Psi}-}\left( 
\begin{array}{c}
p_{x} \\ 
p_{y} \\ 
p_{z}
\end{array}
\right) \left( {\bf \vec{p}\cdot \vec{\Psi}}\right) =0,  \label{n11}
\end{equation}
where ${\bf \vec{\Psi}}$ is a 3 component (column) wave function. Eq. ($\ref
{n11}$) will be satisfied if the two equations

\begin{eqnarray}
\left( \frac{E}{c}I^{\left( 3\right) }+{\bf \vec{p}\cdot \vec{S}}\right) 
{\bf \vec{\Psi}} &=&0,  \label{n12} \\
{\bf \vec{p}\cdot \vec{\Psi}} &=&0,  \label{n13}
\end{eqnarray}
will be simultaneously satisfied. For real energies and momenta complex
conjugation of Eqs. (\ref{n11}) and (\ref{n8}) leads to

\begin{equation}
\left( \frac{E^{2}}{c^{2}}-{\bf \vec{p}}^{2}\right) {\bf \vec{\Psi}}^{\ast }%
{\bf =}\left( \frac{E}{c}I^{\left( 3\right) }+{\bf \vec{p}\cdot \vec{S}}%
\right) \left( \frac{E}{c}I^{\left( 3\right) }-{\bf \vec{p}\cdot \vec{S}}%
\right) {\bf \vec{\Psi}}^{\ast }{\bf -}\left( 
\begin{array}{c}
p_{x} \\ 
p_{y} \\ 
p_{z}
\end{array}
\right) \left( {\bf \vec{p}\cdot \vec{\Psi}}^{\ast }\right) =0,  \label{n13a}
\end{equation}
where ${\bf \vec{\Psi}}^{\ast }$ is the complex conjugate of \ ${\bf \vec{%
\Psi}}.$ Eq. (\ref{n13a}) will be satisfied if the two equations

\begin{eqnarray}
\left( \frac{E}{c}I^{\left( 3\right) }-{\bf \vec{p}\cdot \vec{S}}\right) 
{\bf \vec{\Psi}}^{\ast } &=&0,  \label{n13b} \\
{\bf \vec{p}\cdot \vec{\Psi}}^{\ast } &=&0,  \label{n13c}
\end{eqnarray}
will be simultaneously satisfied. Eqs. (\ref{n11}) and (\ref{n13a}) are the
two different possible decompositions of their left hand side. Eqs. (\ref
{n13a}-\ref{n13c}) do not contain new information as they are only the
complex conjugates of Eqs. (\ref{n11}-\ref{n13}). On the other hand the
physical interpretation is different, namely Eq. (\ref{n12}) is the negative
helicity equation, while Eq. (\ref{n13b}) is the positive helicity equation.
It will be interesting to note that also other set of equivalent equations
is possible. Eqs. (\ref{n11}) and (\ref{n13}) can be rewritten as

\begin{equation}
\left( 
\begin{array}{c}
p_{x} \\ 
p_{y} \\ 
p_{z}
\end{array}
\right) \left( {\bf \vec{p}\cdot \vec{\Psi}}\right) =\left( \frac{E}{c}%
I^{\left( 3\right) }-{\bf \vec{p}\cdot \vec{S}}\right) \left( \frac{E}{c}%
I^{\left( 3\right) }+{\bf \vec{p}\cdot \vec{S}}\right) {\bf \vec{\Psi}-}%
\left( \frac{E^{2}}{c^{2}}-{\bf \vec{p}}^{2}\right) {\bf \vec{\Psi}=0,}
\label{n13d}
\end{equation}
which will be satisfied if the two equations

\begin{equation}
\left( \frac{E}{c}I^{\left( 3\right) }+{\bf \vec{p}\cdot \vec{S}}\right) 
{\bf \vec{\Psi}}=0,\quad \left( \frac{E^{2}}{c^{2}}-{\bf \vec{p}}^{2}\right) 
{\bf \vec{\Psi}=0,}  \label{n13f}
\end{equation}
or their equivalents

\begin{equation}
\left( \frac{E}{c}I^{\left( 3\right) }-{\bf \vec{p}\cdot \vec{S}}\right) 
{\bf \vec{\Psi}}^{\ast }=0,\quad \left( \frac{E^{2}}{c^{2}}-{\bf \vec{p}}%
^{2}\right) {\bf \vec{\Psi}}^{\ast }{\bf =0,}  \label{n13g}
\end{equation}
will be simultaneously satisfied. Maxwell equations will be derived from
Eqs. ($\ref{n12}$-$\ref{n13}$).

We will show below that if in Eqs. ($\ref{n12}$) and ($\ref{n13}$) the
quantum operator substitutions

\begin{equation}
E\Longrightarrow i\hbar \frac{\partial }{\partial t},\quad {\bf \vec{p}%
\Longrightarrow -}i\hbar \nabla ,\quad  \label{n14}
\end{equation}
and the wavefunction substitution

\begin{equation}
{\bf \vec{\Psi}=\vec{E}-}i{\bf \vec{B},}  \label{n15}
\end{equation}
are made, as a result the Maxwell equations will be obtained. In Eq. (\ref
{n15}) ${\bf \vec{E}}$ and ${\bf \vec{B}}$ are the electric and magnetic
fields respectively. Indeed, one can easily check from Eqs. (\ref{n8}) and (%
\ref{n14}) that the following identity is satisfied

\begin{equation}
\left( {\bf \vec{p}\cdot \vec{S}}\right) {\bf \vec{\Psi}=}\hbar \nabla
\times {\bf \vec{\Psi}.}  \label{n16}
\end{equation}
From Eqs. (\ref{n12}-\ref{n13}) and (\ref{n14}, \ref{n16}) we obtain

\begin{equation}
\frac{i\hbar }{c}\frac{\partial }{\partial t}{\bf \vec{\Psi}=}-\hbar {\bf %
\nabla }\times {\bf \vec{\Psi},}  \label{n17}
\end{equation}

\begin{equation}
-i\hslash \nabla \cdot {\bf \vec{\Psi}=}0.  \label{n18}
\end{equation}
The constant $\hslash $ can be cancelled out in Eqs. (\ref{n17}-\ref{n18}),
and after replacing ${\bf \vec{\Psi}}$ by Eq. (\ref{n15}), the following
equations are obtained

\begin{equation}
{\bf \nabla }\times \left( {\bf \vec{E}-}i{\bf \vec{B}}\right) {\bf =-}i%
\frac{1}{c}\frac{\partial \left( {\bf \vec{E}-}i{\bf \vec{B}}\right) }{%
\partial t},  \label{n19}
\end{equation}

\begin{equation}
{\bf \nabla }\cdot \left( {\bf \vec{E}-}i{\bf \vec{B}}\right) {\bf =}0.
\label{n20}
\end{equation}
If in Eqs. (\ref{n19}-\ref{n20}) the electric and magnetic fields are real,
the separation into the real and imaginary parts will lead to the Maxwell
equations

\begin{equation}
{\bf \nabla }\times {\bf \vec{E}=-}\frac{1}{c}\frac{\partial {\bf \vec{B}}}{%
\partial t}  \label{1}
\end{equation}

\begin{equation}
{\bf \nabla }\times {\bf \vec{B}=}\frac{1}{c}\frac{\partial {\bf \vec{E}}}{%
\partial t}  \label{2}
\end{equation}

\begin{equation}
{\bf \nabla }\cdot {\bf \vec{E}}=0  \label{3}
\end{equation}

\begin{equation}
{\bf \nabla }\cdot {\bf \vec{B}}=0.  \label{4}
\end{equation}

One should note that the Plank constant $\hslash $ was cancelled out
earlier, in Eqs. (\ref{n17}-\ref{n18}), which explains its absence in the
Maxwell equations. Another comment should be made here, starting from
equations (\ref{n13f}) and (\ref{n14}-\ref{n15}) one can get equations which
are equivalent to Maxwell equations (without sources), namely

\begin{equation}
{\bf \nabla }\times {\bf \vec{E}=-}\frac{1}{c}\frac{\partial {\bf \vec{B}}}{%
\partial t},\quad {\bf \nabla }\times {\bf \vec{B}=}\frac{1}{c}\frac{%
\partial {\bf \vec{E}}}{\partial t},\quad \left( \frac{\partial ^{2}}{%
c^{2}\partial t^{2}}-{\bf \nabla }^{2}\right) \overrightarrow{{\bf E}}%
=0,\quad \left( \frac{\partial ^{2}}{c^{2}\partial t^{2}}-{\bf \nabla }%
^{2}\right) \overrightarrow{{\bf B}}=0,  \label{4a}
\end{equation}
the Gauss laws (\ref{3}-\ref{4}) are satisfied on the basis of Eq. (\ref
{n13d}).

Dirac \cite{dirac} and Wigner \cite{wigner},\cite{bacry} have derived
relativistic equations for massless particles of any spin from which the
Gauss laws, for the spin one case, can be derived. Moreover, Wigner \cite
{wigner} has shown that any finite-component massless field has only two
possible helicity states. Dirac has derived equations for massless particles
with spin $k$, which in the ordinary vector notation \cite{dirac} are

\begin{equation}
\left\{ kp_{t}+S_{x}p_{x}+S_{y}p_{y}+S_{z}p_{z}\right\} \psi =0,  \label{a1}
\end{equation}

\begin{equation}
\left\{ kp_{x}+S_{x}p_{t}-iS_{y}p_{z}+iS_{z}p_{y}\right\} \psi =0,
\label{a2}
\end{equation}

\begin{equation}
\left\{ kp_{y}+S_{y}p_{t}-iS_{z}p_{x}+iS_{x}p_{z}\right\} \psi =0,
\label{a3}
\end{equation}

\begin{equation}
\left\{ kp_{z}+S_{z}p_{t}-iS_{x}p_{y}+iS_{y}p_{x}\right\} \psi =0,
\label{a4}
\end{equation}
where the $p_{n}$ are the momenta, $p_{t}=E/c$, $E$ the energy, $\psi $ a $%
\left( 2k+1\right) $ component wave function and $S_{n}$ are the spin $%
\left( 2k+1\right) \times \left( 2k+1\right) $ matrices which satisfy

\begin{equation}
\left[ S_{x},S_{y}\right] =iS_{z},\quad \left[ S_{z},S_{x}\right]
=iS_{y},\quad \left[ S_{y},S_{z}\right] =iS_{x},\quad
S_{x}^{2}+S_{y}^{2}+S_{z}^{2}=k(k+1)I^{\left( k\right) },  \label{a5}
\end{equation}
and $I^{\left( k\right) }$ is a $\left( 2k+1\right) \times \left(
2k+1\right) $ unit matrix. As we shall see below, for the case $k=1$, Eq. (%
\ref{a1}) will lead to the Faraday and Ampere-Maxwell laws. The Gauss laws
can be derived from Eqs. (\ref{a1}-\ref{a4}) in a way which will be
described below. Eqs. (\ref{a1}-\ref{a4}) were analyzed extensively by Bacry 
\cite{bacry}, who derived them using Wigner's condition \cite{wigner} on the
Pauli-Lubanski vector $W^{\mu }$ for massless fields

\begin{equation}
W^{\mu }=kp^{\mu },\quad \mu =x,y,z,t.  \label{a6}
\end{equation}

\bigskip Let us now demonstrate how Eq. (\ref{n13}) can be derived from Eqs.
(\ref{a1}-\ref{a4}). Following Dirac \cite{dirac}, one replaces Eqs. (\ref
{a1}-\ref{a4}), which are linearly dependent, with the Eq. (\ref{a1}) and 3
conditions on the wave function, which are obtained by substituting $p_{t}$
from Eq. (\ref{a1}) into Eqs. (\ref{a2}-\ref{a4})

\begin{equation}
\left\{ kp_{t}+S_{x}p_{x}+S_{y}p_{y}+S_{z}p_{z}\right\} \psi =0,  \label{b1}
\end{equation}

\begin{equation}
\left\{ (k^{2}-S_{x}^{2})p_{x}+(ikS_{z}-S_{x}S_{y})p_{y}-\left(
ikS_{y}+S_{x}S_{z}\right) p_{z}\right\} \psi =0,  \label{b2}
\end{equation}

\begin{equation}
\left\{ (k^{2}-S_{y}^{2})p_{y}+(ikS_{x}-S_{y}S_{z})p_{z}-\left(
ikS_{z}+S_{y}S_{x}\right) p_{x}\right\} \psi =0,  \label{b3}
\end{equation}

\begin{equation}
\left\{ (k^{2}-S_{z}^{2})p_{z}+(ikS_{y}-S_{z}S_{x})p_{x}-\left(
ikS_{x}+S_{z}S_{y}\right) p_{y}\right\} \psi =0.  \label{b5}
\end{equation}
For the case $k=1$, $\psi \equiv \vec{\Psi}$, and using the representation (%
\ref{n8}) for the spin matrices, one obtains for Eq. (\ref{b2})

\begin{equation}
\left( 
\begin{array}{ccc}
p_{x} & p_{y} & p_{z} \\ 
0 & 0 & 0 \\ 
0 & 0 & 0
\end{array}
\right) \vec{\Psi}=0,  \label{b6}
\end{equation}
for Eq. (\ref{b3})

\begin{equation}
\left( 
\begin{array}{ccc}
0 & 0 & 0 \\ 
p_{x} & p_{y} & p_{z} \\ 
0 & 0 & 0
\end{array}
\right) \vec{\Psi}=0,  \label{b7}
\end{equation}
and for Eq. (\ref{b5})

\begin{equation}
\left( 
\begin{array}{ccc}
0 & 0 & 0 \\ 
0 & 0 & 0 \\ 
p_{x} & p_{y} & p_{z}
\end{array}
\right) \vec{\Psi}=0,  \label{b8}
\end{equation}
which all are equivalent to Eq. (\ref{n13}). It is interesting to note that

\begin{eqnarray}
\left( 
\begin{array}{ccc}
p_{x}^{2} & p_{x}p_{y} & p_{x}p_{z} \\ 
p_{y}p_{x} & p_{y}^{2} & p_{y}p_{z} \\ 
p_{z}p_{x} & p_{z}p_{y} & p_{z}^{2}
\end{array}
\right) &=&\left( 
\begin{array}{ccc}
p_{x} & 0 & 0 \\ 
p_{y} & 0 & 0 \\ 
p_{z} & 0 & 0
\end{array}
\right) \left( 
\begin{array}{ccc}
p_{x} & p_{y} & p_{z} \\ 
0 & 0 & 0 \\ 
0 & 0 & 0
\end{array}
\right)  \label{b9} \\
&=&\left( 
\begin{array}{ccc}
0 & p_{x} & 0 \\ 
0 & p_{y} & 0 \\ 
0 & p_{z} & 0
\end{array}
\right) \left( 
\begin{array}{ccc}
0 & 0 & 0 \\ 
p_{x} & p_{y} & p_{z} \\ 
0 & 0 & 0
\end{array}
\right)  \label{b10} \\
&=&\left( 
\begin{array}{ccc}
0 & 0 & p_{x} \\ 
0 & 0 & p_{y} \\ 
0 & 0 & p_{z}
\end{array}
\right) \left( 
\begin{array}{ccc}
0 & 0 & 0 \\ 
0 & 0 & 0 \\ 
p_{x} & p_{y} & p_{z}
\end{array}
\right) ,  \label{b11}
\end{eqnarray}
from which we deduce that the equations (\ref{n12}) and (\ref{n13}) and the
decomposition (\ref{n7}), can be realized on the basis of Eq. (\ref{b1}) and
one of the equations (\ref{b2}),(\ref{b3}) and (\ref{b5}).

Above, we have shown how all Maxwell equations can be derived simultaneously
from first principles, similar to those which have been used to derive the
Dirac relativistic electron equation. Moreover the wave function ${\bf \vec{%
\Psi}}$ has a definite local classical interpretation in terms of the
electric and magnetic fields, as given by Eq. (\ref{n15}), which are locally
measurable and well understood quantities. Therefore Maxwell equations
should be used as a guideline for proper interpretations of quantum theories.

\end{document}